\def\year{2016}
\documentclass[letterpaper]{article}
\usepackage{aaai}
\usepackage{times}
\usepackage{helvet}
\usepackage{hhline}
\usepackage{courier}

\usepackage{balance}  % to better equalize the last page
\usepackage{graphics} % for EPS, load graphicx instead 
\usepackage{txfonts}
\usepackage{times}    % comment if you want LaTeX's default font
\usepackage[pdftex]{hyperref}
\usepackage{color}
\usepackage{textcomp}
\usepackage{booktabs}
\usepackage{ccicons}
\usepackage{todonotes}
\usepackage{xcolor}

\usepackage{helvet}
\usepackage{courier}
\usepackage{times}
\usepackage{url}
\usepackage{latexsym}
\usepackage{graphicx, color, soul}
\usepackage{balance}
\usepackage[font={small}]{caption}
\usepackage[labelfont=bf]{caption}
\usepackage[]{algorithm2e}
\usepackage{subcaption}
\usepackage{bbm}
\nocopyright

\begin{document}

\title{Trending Chic: Analyzing the Influence of Social Media on Fashion Brands}
\author{Lydia Manikonda \qquad Ragav Venkatesan \qquad Subbarao Kambhampati \qquad Baoxin Li \\ Department of Computer Science, Arizona State University, Tempe AZ 85281\\
\{lmanikon, ragav.venkatesan, rao, baoxin.li\}@asu.edu
}

\maketitle

\begin{abstract}
Social media platforms are popular venues for fashion brand marketing and advertising. With the introduction of native advertising, users don't have to endure banner ads that hold very little saliency and are unattractive. Using images and subtle text overlays, even in a world of ever-depreciating attention span, brands can retain their audience and have a capacious creative potential. While an assortment of marketing strategies are conjectured, the subtle distinctions between various types of marketing strategies remain under-explored. This paper presents a qualitative analysis on the influence of social media platforms on different behaviors of fashion brand marketing. We employ both linguistic and computer vision techniques while comparing and contrasting strategic idiosyncrasies. We also analyze brand audience retention and social engagement hence providing suggestions in adapting advertising and marketing strategies over Twitter and Instagram.
\end{abstract}

\section{Introduction}
\label{sec:intro}

The impact of fashion in society has been a well-studied topic even during the era of print and visual media based advertising~\cite{fashion_impact1,fashion_impact2}. The marketing and advertising strategies involved in fashion are often qualitatively different from other product marketing and advertising. While in most products it is important to emphasize the necessity or the quality of the product, fashion advertisements are tailor-made to match the tastes and sensibilities of the target audience. 

Social media is an an incredible tool at the fashion industry's disposal for marketing. By leveraging social media, brands can take control of public perception which is one among the many important factors in fashion marketing~\cite{customer}. The continuous feedback received by the brands via the likes and comments on their social media posts lets them gauge and further viralize their base in the market.
There are several studies that focus on understanding the growing interest in social media marketing~\cite{mark:luxury,Kim:social,ang:ko}. The importance of fashion branding on social media is becoming even more pronounced as networks like Instagram are revolutionizing this field. According to the well-analyzed editorials in The Guardian and The New York Times, it is the social media that decides what you wear and  Instagram is titled as the \emph{fashion's new front row}~\cite{fashion:insta2,fashion:insta4,fashion:insta3}. Existing literature~\cite{yuheng2014} shows that Instagram alone has a significant share of posts that belong to fashion category. 

In this paper, we consider the top-$20$ fashion brands and investigate how they use Twitter and Instagram by observing their native profiles. We analyze their styles and strategies of advertisement. Although textual analysis is interesting, we predominantly focus on visual analysis owing to the overwhelming number of images used by the brands in advertising. Using the neural network based deep image features similar to those extracted by Khosla et al.~\cite{khosla2014makes},  we find out how the two types of brand marketing strategies -- \textit{direct marketing} and \textit{indirect marketing} are used.  Our analysis revealed that brands that have a larger number of visibility tend to utilize the direct marketing strategy. 

The summary of our contributions is as follows:
\begin{itemize}
\item A characterization of how top-20 fashion brands use the social media primarily to compare and contrast the posts on Twitter and Instagram.
\item Using deep features obtained from the visual content, an investigation about two marketing strategies -- direct marketing and indirect marketing. 
\end{itemize}

We hope that the distinctions discovered in this paper can inspire marketing researchers to study the reason behind these inferences. To the best of our knowledge, there is no existing work on how fashion brands use different social platforms in terms of characterizing their behavior through analyzing textual and visual content. It is important to understand the distinctions and similarities so that the new businesses can adapt these ideas to promote their businesses and establish their brands. 
%%%%%%%%%%%%%%%%%%%%%%%%%%%%%%%%%%%%%%%%%%%%%%%%%%%%%%%%%%%%%%%%%%%%%%%%%%%%%%%5
%section{Dataset}
\section{Analysis}

The dataset used in this analysis comprises of top-$20$ fashion brands (in terms of the number of followers) on Instagram and Twitter ($D=\{b_1, b_2, ... b_{20}\}$ in Table~\ref{tab:diffbrands}) according to the survey conducted by Harper's Bazaar~\cite{harper:survey}. Harper's Bazaar is a monthly fashion magazine that delivers a perspective into the world of fashion, beauty and popular culture and is considered as a good style resource for women. 
%Consider the dataset as containing brands $D=\{b_1, b_2, ... b_{20}\}$ where $b_1$, $\dots$, $b_{20}$ are the brands that we consider for analysis on both Twitter and Instagram.

\begin{table}
\small
%\vspace{-2mm}
\centering
\caption{Top-20 brands used in this study}
\begin{tabular}{|c|c|} \hline
Nike -- $b_1$ & Adidas Originals (AO) -- $b_2$ \\ \hline
Louis Vuitton (LV) -- $b_3$ & Dolce Gabbana (DG) -- $b_4$ \\ \hline
Michael Kors (MK) -- $b_5$ & Adidas -- $b_6$ \\ \hline
Dior -- $b_7$ & Louboutin World (LW) -- $b_8$ \\ \hline
Gucci -- $b_9$ & Prada -- $b_{10}$ \\ \hline
Burberry (Brb) -- $b_{11}$ & Vans -- $b_{12}$ \\ \hline
Fendi -- $b_{13}$ & Armani -- $b_{14}$ \\ \hline
Converse -- $b_{15}$ & Jimmy Choo (JC) -- $b_{16}$ \\ \hline
Free People (FP)  -- $b_{17}$ & Calvin Klein (CK) -- $b_{18}$ \\ \hline
Ralph Lauren (RL) -- $b_{19}$ & Cartier -- $b_{20}$ \\ \hline
\end{tabular}
%\vspace{-2mm}
\label{tab:diffbrands}
\end{table}

%To download the tweets and meta information associated with each tweet for these brands in $D$ on Twitter, we use the Twython API~\footnote{\url{https://twython.readthedocs.org/en/latest/}}. We downloaded all the tweets including metadata with respect to followers, friends and media posted. On Instagram, we used the programming API~\footnote{\url{http://instagram.com/developer/}} to collect the data for brands present in $D$ that includes photos along with the meta data associated with the photo viz., number of likes, comments, caption, geolocation and hashtags. 
\subsection{Group statistics}
\begin{figure*}
%\tiny
\small
\centering
\includegraphics[width=\linewidth]{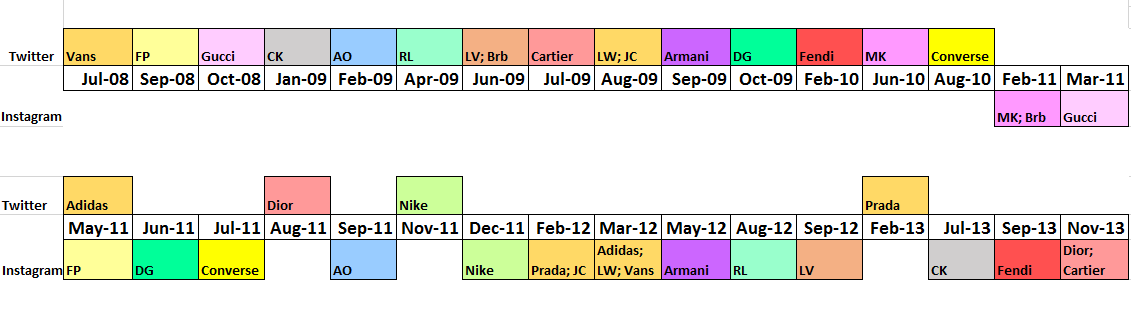}
\caption{A timeline showing the creation dates of accounts by the brands on Twitter and Instagram}
\label{fig:timeline}
\end{figure*}

Since Twitter was founded in 2006 and Instagram in 2010, we see in Figure~\ref{fig:timeline} that most brands had their accounts created on Twitter first. Figure~\ref{fig:timeline} shows the timeline of when the brands created their accounts and first posted a tweet or photo on Twitter or Instagram respectively. Among all these brands the first post was made by Vans in 2008 on Twitter and by Michael Kors in 2011 on Instagram. Every month (on an average) a minimum of $18$ posts and a maximum of $124$ posts on Instagram and a minimum of $20$ posts and a maximum of $619$ posts on Twitter were made by these brands. 

\begin{figure*}
\tiny
\vspace{-2mm}
\centering
\includegraphics[width=0.49\linewidth]{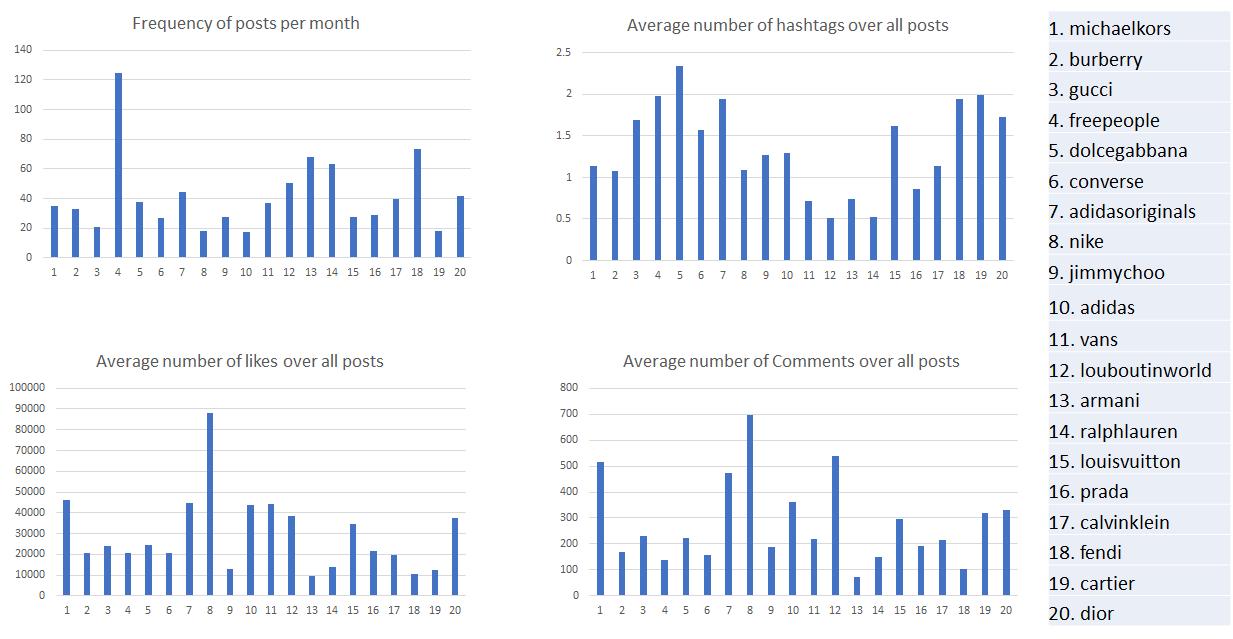}  
~
\includegraphics[width=0.49\linewidth]{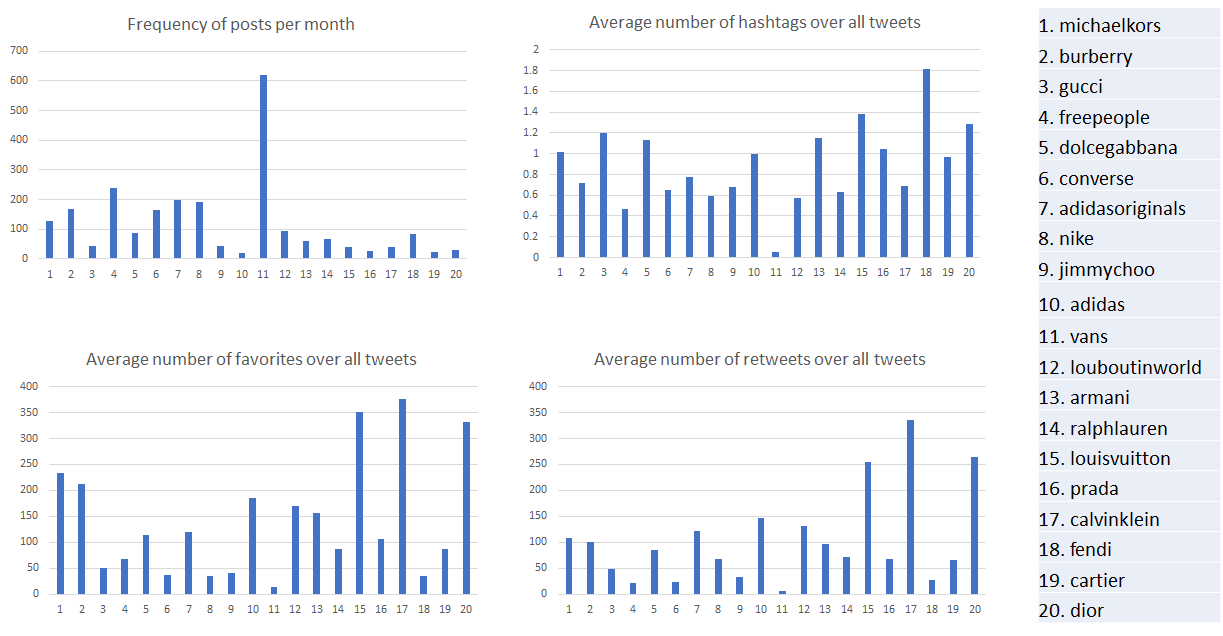}
\caption{Different statistics showing the brand behavior on Instagram and twitter}
\label{fig:stats}
\vspace{-2mm}
\end{figure*}

%%%%%%%%%%%%%%%%%%%%%%%%%%%%%%%%%%%%%%%%%%%%%%%%%%%%%%%%%%%%%%%%%%%%%%%%%%%%%%%%%%%%%%%%%
The frequency of posts per month by each brand using eq.~\ref{eq:calcfreq} is computed along with the average number of likes, average number of comments and the average number of hashtags for all posts of brands~\ref{fig:stats}. Suppose $n _j^{(t,b)}$ and $n _j^{(i,b)}$ are the number of posts made on Twitter and Instagram by the brand $b$ respectively in the time period $j$\footnote{For convenience purpose, we use a monthly time period}. $p = \{t, i\}$ refer to Twitter and Instagram respectively,

\begin{equation}
%M_i^{(p)} = \mathcal{I}\{ \sum m_i^{(p)} \} \forall i,
M_i^{(p)} = \mathbbm{1}\{ \sum m_i^{(p)} \} \forall i,
\end{equation}
is the vector that is an indicator function ($\mathbbm{1}$) for each time period, indicating whether any posts at all were made during that time period. It is $1$ if any posts were made and $0$ if not. Then,

\begin{equation}
\omega_p = \frac{P_t}{\sum M_i^{(p)}}, 
\label{eq:calcfreq}
\end{equation}

is the average frequency of posts made by a brand on the platform $p$.

%avgfreq = \frac{Total\ number\ of\ posts }{Total\ months\ where\ atleast\ one\ post\ was\ made}
%\end{equation}

From Figure~\ref{fig:stats}, we notice that the brands Michael Kors (MK) and Burberry (Brb) created their accounts at the same time and have similar number of posts. But in terms of the number of followers, MK has $24\%$ more number of followers than Brb, follows twice the number of people followed by Brband gets $3$ times as many likes and comments than Brb. Brands like Free People (FP) created an Instagram account during summer of 2011 and post pictures with a very high frequency (124 pictures on average every month). When we observe the number of likes and comments, they are around $20k$ and $136$ respectively which are neither high nor low compared to other brands. These examples suggest that the rate at which a brand makes posts on these platforms do not have any effect on the visibility of posts. 

Users can make posts on Twitter in two forms: (a) submissions that are uploaded to and hosted on the Twitter's server itself and (b) cross-shared posts where tweets posted from another application appear on Twitter timeline. On an average, $87\%$ of tweets contain pictures of the products or models showcasing the brand's products. 

For all the brands on the two platforms in a head-to-head comparison on the number of posts, followers and friends, we find some surprising insights through Figure~\ref{fig:stats}. The key observation is that the posts made on Instagram get very high visibility in terms of the large number of likes compared to Twitter. We associate this with two patterns we observed in our analysis $1)$ People tend to like images more than text posts and Instagram allows only texts $2)$ Instagram doesn't allow cross-sharing unless an image exists, while Twitter allows and followers may not be receptive of posts that involve web urls. It is also a noticeable fact that most of these brands have more number of followers on Instagram than on Twitter.

\paragraph{Popular Trends or topics:} We use the LDA approach~\cite{Blei:LDA} to understand how the brands focus on different topics using the textual features. To discover the topics, we use the Twitter LDA package~\cite{Zhao2011twit}. We consider the captions (for Instagram posts) or tweet text (for Twitter posts) attached with all the posts of a brand and mine the topics across all the brands on the both the platforms. Using LDA, we found 10 topics across all the brands on both the platforms. Table~\ref{tab:topicidwords} presents the 10 words associated with the discovered topics along with the brands which used the topic ID in the same corresponding row on both Twitter and Instagram. 

\begin{table}[ht]
\tiny
\centering
\caption{Topic IDs and their corresponding words}
\begin{tabular}{|p{1mm}|p{5cm}|p{2cm}|} \hline
%\begin{tabular}{|p|p|p|} \hline
\textbf{ID} & \textbf{Words} & \textbf{Brands}\\ \hline
\textbf{0} & \textit{red}, \textit{contact}, \textit{make}, \textit{pack}, \textit{collection}, \textit{team}, \textit{hit}, \textit{online}, \textit{time}, \textit{stores} & Louboutin World, Cartier\\ \hline
\textbf{1} & \textit{show}, \textit{louis}, \textit{fashion}, \textit{vuitton}, \textit{men's}, \textit{collection}, \textit{gucci}, \textit{watch}, \textit{opening}, \textit{live} & Louis Vuitton\\ \hline
\textbf{2} & \textit{collection}, \textit{bag}, \textit{discover}, \textit{show}, \textit{shoeoftheday}, \textit{style}, \textit{botique}, \textit{fashion}, \textit{wearing}, \textit{watch} & Gucci, Fendi\\ \hline
\textbf{3} & \textit{show}, \textit{live}, \textit{personalized}, \textit{moment}, \textit{autumn/winter}, \textit{runway}, \textit{wearing},  \textit{collection}, \textit{british}, \textit{london} & Burberry\\ \hline
\textbf{4} & \textit{win}, \textit{photo}, \textit{pair}, \textit{signed}, \textit{metro}, \textit{entered}, \textit{sean}, \textit{big}, \textit{attitudes}, \textit{submission} & \\ \hline
\textbf{5} & \textit{armani}, \textit{giorgio}, \textit{wearing}, \textit{show}, \textit{fashion}, \textit{celebs}, \textit{summer}, \textit{emporio}, \textit{collection}, \textit{awards} & Dolce Gabbana, Prada, Armani\\ \hline
\textbf{6} & \textit{rl}, \textit{collection}, \textit{regram}, \textit{polo}, \textit{vans}, \textit{rad}, \textit{photo}, \textit{mix}, \textit{fall}, \textit{hope} & Ralph Lauren\\ \hline
\textbf{7} & \textit{styletip}, \textit{conditions}, \textit{merci}, \textit{accessories}, \textit{$générales$}, \textit{live}, \textit{peux-tu}, \textit{participation}, \textit{jetsetgo}, \textit{timeless} & Michael Kors, Adidas\\ \hline
\textbf{8} & \textit{fashion}, \textit{blog}, \textit{love}, \textit{streetstyle}, \textit{photo}, \textit{today}, \textit{fashionista}, \textit{happy}, \textit{inspiration}, \textit{fashionphotography} & \\ \hline
\textbf{9} & \textit{collection}, \textit{show}, \textit{spring}, \textit{runway}, \textit{fall}, \textit{discover},\textit{fashion}, \textit{live}, \textit{dress}, \textit{backstage} & Dior, Calvin Klein\\ \hline
\end{tabular}
\vspace{-2mm}
\label{tab:topicidwords}
\end{table}

We can notice that the runway luxury fashion brands like Louis Vuitton, Dolce Gabbana, Burberry, etc., focus on the same topics in Twitter and Instagram. Very few brands like Nike, Adidas Originals, Vans, Converse and Free People focus on different topics on the two platforms. Nike being the most popular brand on Instagram focuses mainly on topic 7 whereas on Twitter focuses on topic 0 whose words suggest that Twitter might be used for correspondence or queries. We find that brands like Burberry focus mainly on British style and includes men's collections and Michael Kors focuses on the styles and accessories. The active brand according to the number of friends -- Louboutin World uses both the networks to contact customers online.

\subsection{Visual Features}

%Deep learning is being used to convert data of various modalities into representations where the entropy is structured and ordered according to some tasks they are trained for. 
In analyzing the visual content on both the social networks, we use the dataset of images collected from the brand accounts on Twitter and Instagram and we extract deep features for each image present in our dataset. In this article, we use the overfeat networks' image features~\cite{sermanet-iclr-14} for two reasons. As argued by Khosla et al., overfeat-type features are particularly capable of extracting representations that are well-suited for internet images and abstract tasks. Overfeat is a stable implementation that makes use of GPU (we used Tesla K40) in the efficient extraction of features for large scale image datasets. 

\begin{figure*}[t]
\tiny
\vspace{-2mm}
\centering
\begin{subfigure}[b]{0.325\textwidth}
\includegraphics[width=\textwidth]{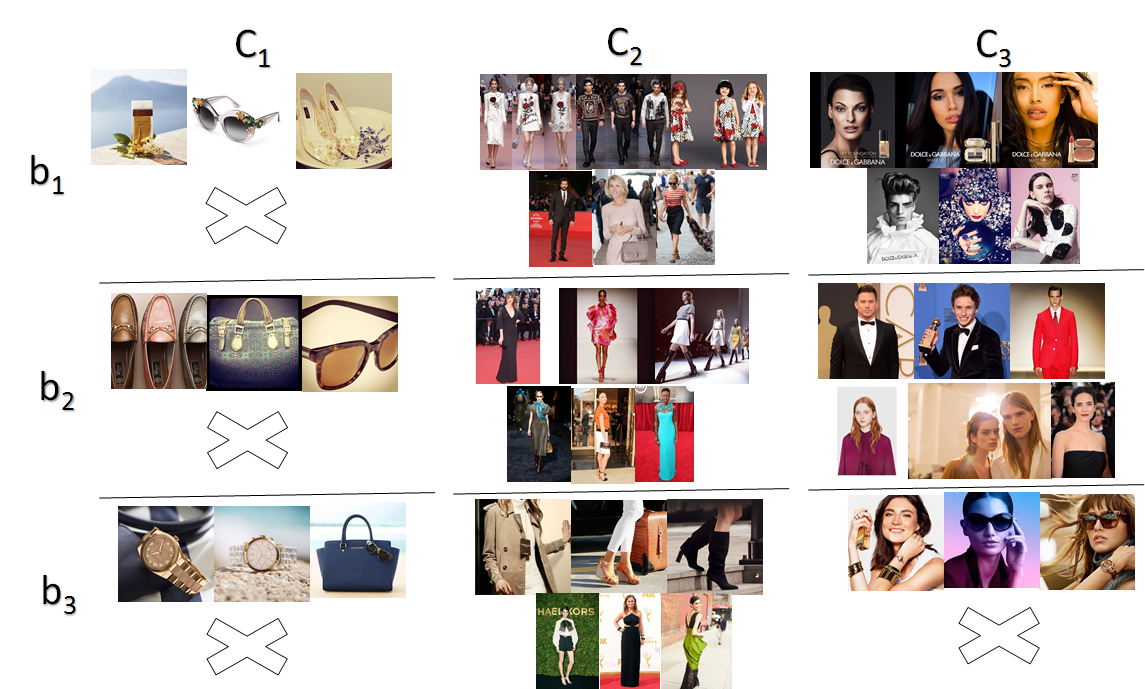} \vspace{-2mm} \caption{} ~
\end{subfigure}
\begin{subfigure}[b]{0.325\textwidth}
\includegraphics[width=\textwidth]{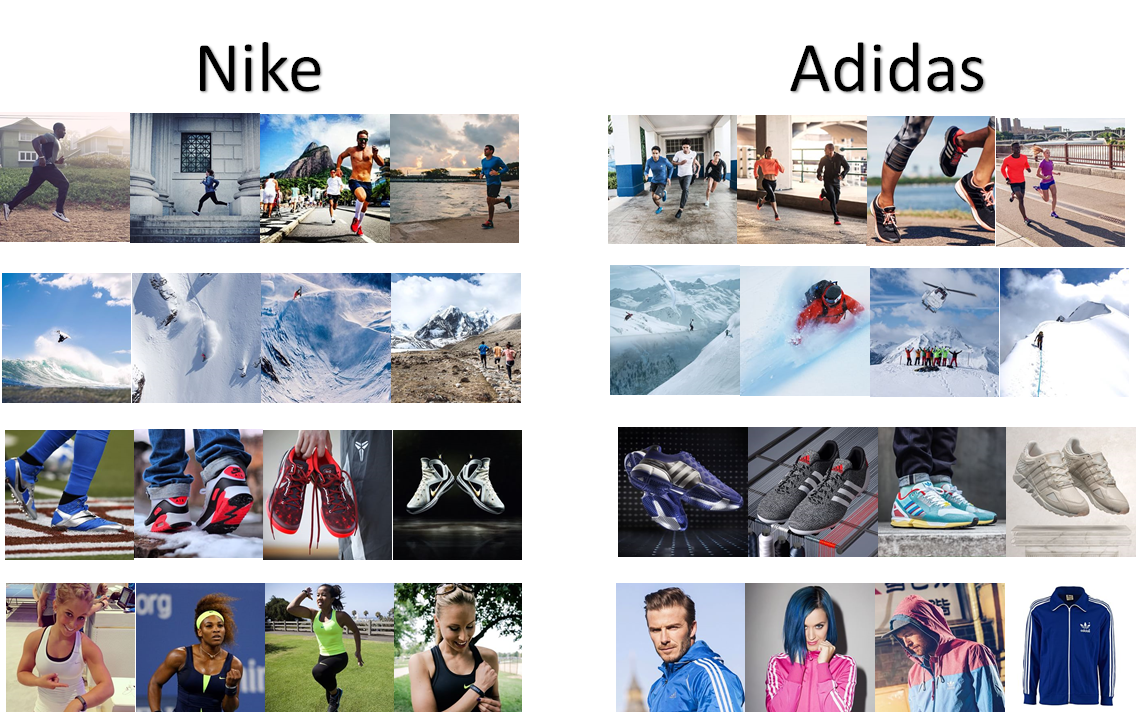} \vspace{-2mm} \caption{} ~
\end{subfigure}
\begin{subfigure}[b]{0.325\textwidth}
\includegraphics[width=\textwidth]{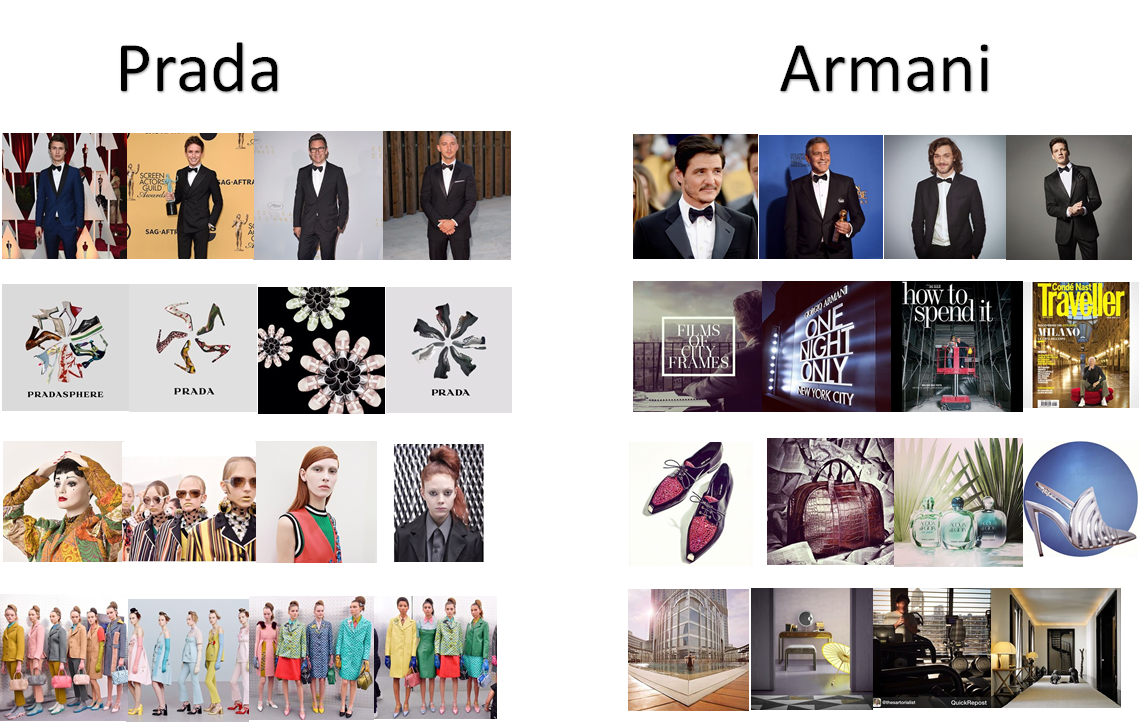} \vspace{-2mm} \caption{}
\end{subfigure}
\caption{a) Michael Kors vs Gucci vs D\& G, b) Nike vs Adidas, c) Armani vs Prada}
\label{fig:visual_comparison}
\vspace{-4mm}
\end{figure*}

We use the network's $22^{\textit{nd}}$ layer representation for each image as the feature vector corresponding to that image. We then perform clustering (using $k$-means) on this space and use those clusters to study the different marketing strategies utilized by the brands and how it affects the visibility of their products. We obtain $k$ clusters for each brand on the two platforms separately. These clusters represent the different types of content categories present in the images for example -- sunglasses, watches, floral patterns, etc. Figure~\ref{fig:visual_comparison} (a) indicates that brands which post similar textual topics across these platforms post different types of visual content. 

%For instance, if a deep Convolutional Neural Network is trained for some prosaic task such as object recognition, they learn a distributed representation and map the images on to a vector space that carries details related to arbitrary semantics such as objectness, color, texture etc. Depending on the layer at which these representations are probed, one can glean a reasonable semantics. For instance, the earlier layers of a deep CNN seem to encode more edge and Gabor-like features while the latter layers seem to encode more meaningful features. It is a common practice to use these \emph{off-the-shelf} pre-learned networks on other tasks as imaging feature extractors. 

We identified two distinct and common strategies that brands use -- direct product marketing (DM) and indirect product marketing (IM). DM focuses on the product and IM uses attributes that are not but related to the product for marketing. For instance, a bag that is photographed by itself on a pedestal is DM, while a fashion model (person) holding the bag and the bag being vignetted is an example of IM. We often find both, while we find that IM is more effective, particularly when used with celebrities. The same can be observed in the following cluster analysis over the said feature space. Figure~\ref{fig:visual_comparison} displays the brands $b_1$ -- Dolce Gabbana, $b_2$ -- Gucci, $b_3$ -- Michael Kors along the cluster types -- $C_1$ -- Products, $C_2$ -- Runway/Redcarpet events, $C_3$ -- Portraits for both Instagram (top row) and Twitter (bottom row). Brand $b_1$ focuses on direct marketing on Instagram but doesn't make posts of category $C_1$ where as it focuses on indirect marketing w.r.t category $C_3$. Brand $b_2$ follows the similar trend as $b_1$. Whereas, brand $b_3$ primarily focuses on indirect marketing no matter what category it is. While $b_1$ and $b_2$ have mean likes of $27245$ and $25280$ respectively, brand $b_3$ has $47941$ likes on average for the said clusters. Similar pattern is spread across many other similar brands, strongly favoring indirect marketing and aligning with the past research~\cite{faces:CHI}.

Among all the brands, Nike has the largest number of followers and the number of likes received for a post. Adidas focuses on the same topics as Nike giving us a good case study. Each row in Figure~\ref{fig:visual_comparison} corresponds to a cluster category where Nike and Adidas both post similar kinds of photos except that Nike and Adidas has a unique cluster focusing on the tank tops and track jackets respectively. Both the brands focus on direct and indirect marketing in very similar patterns. Direct marketing for shoes and indirect marketing for equipment and attire. We notice that Nike and Adidas acquire similar number of likes for most similar categories. On the idiosyncratic categories we find that Nike posting tank tops get significantly larger number of likes than the track-suits of Adidas. Nike also gets significantly more likes due to the presence of their Tennis celebrities Rafael Nadal and Serena Williams in Instagram and this is the major cause of Nike having more likes and followers than Adidas. 

We extend similar analysis to two runway brands Prada and Armani, which focus on similar topics. Figure~\ref{fig:visual_comparison} shows that even if both the brands has some common clusters, there are distinctive cluster categories. Prada posts often contain floral patterns with no architecture and not much focused on products. Where as, Armani has photos with text, photos that focus on indoor architecture and photos of products. Prada having developed significant following for is floral pattern earns four times as many likes and  comments for that cluster as compared to the comparable cluster in Armani. While the men in tuxes in both brands earn similar number of likes and comments, the architecture cluster of Armani gets significantly less number of likes and comments. 

Following the above discussed trends, we find time and again that posts made by brands practicing IM strategies have more visibility. We hope that these explorations could draw the attention of market researchers that IM leads to more visibility in terms of obtaining more likes and comments for posts on Instagram. 

%\vspace{-2mm}
\section{Conclusions}
Our work employs linguistic and visual analyses on the posts made by top-20 fashion brands on Twitter and Instagram. We investigate how brands focus on different topics on different social media and how certain types of visual cues associated with marketing strategies can obtain more visibility. Textual analysis revealed that in spite of the number of hashtags a post contains or how frequently a brand makes posts online, do not contribute to visibility. Visual analyses show that even if the textual topics are same on both the platforms, brands adapt different posting styles w.r.t visual content. However, it was evident from the analysis that brands exercising indirect marketing are gaining more visibility in terms of the number of likes and comments. Through this research we hope to open up new discussions about the role of visual content on social media in learning about fashion trends and inspire marketing researchers to study the reasons behind these findings. 

%\vspace{-4mm}
\small
\bibliographystyle{aaai}
\bibliography{refs}

\end{document}